\documentstyle[aps]{revtex}
\begin{document}
\draft
\preprint{UMP--96--83}
\title{\large\bf 
2D XXZ Model ground state Properties using an analytic Lanczos Expansion}
\author{\large 
N.S. Witte$^{1}$\footnote{E-mail: {\tt nsw@physics.unimelb.edu.au}},
Lloyd C.L. Hollenberg$^{1}$ 
and Zheng Weihong$^{2}$}
\address{
$^{1}$ Research Centre for High Energy Physics, \\
School of Physics, University of Melbourne,\\
Parkville, Victoria 3052, AUSTRALIA.}
\address{
$^{2}$ School of Physics, University of New South Wales,\\
Sydney, NSW 2052, AUSTRALIA.}
\maketitle
\begin{abstract}
We develop the formalism for calculating arbitrary expectation values 
for any extensive lattice Hamiltonian system
using a new analytic Lanczos expansion, or plaquette expansion, and
a recently proved exact theorem for ground state energies.
The ground state energy, staggered magnetisation and the excited state
gap of the 2D anisotropic antiferromagnetic Heisenberg Model are then 
calculated using this 
expansion for a range of anisotropy parameters and compared to other
moment based techniques, such as the {\it t}-expansion, and spin-wave
theory and series expansion methods.
We find that far from the isotropic point all moment methods give
essentially very similar results, but near the isotropic point the
plaquette expansion is generally better than the others.
\end{abstract}

\pacs {PACS: 11.15.Tk, 05.30.-d, 75.10.Jm, 05.50.+q}
\eject
\section{Introduction}
In this work we employ a very recent addition to the range of 
moment methods available now to the problem of the $ s=1/2 $ 
antiferromagnetic Heisenberg model on a square lattice.
These other methods are based essentially on linked-cluster expansions
and include the {\it t}-expansion\cite{hw84}, 
the connected moment expansion (CMX)\cite{c87},
and the coupled-cluster expansion\cite{bk87},\cite{b91}.
The method we use is based on the Lanczos method, 
\begin{equation}
  \hat{H} |\psi_n\rangle =
          \beta_n |\psi_{n-1}\rangle 
                  + \alpha_n |\psi_n\rangle 
                             + \beta_{n+1}|\psi_{n+1}\rangle \ ,
\end{equation}
starting with a trial state $ |\psi_0\rangle $,
whereby approximate yet analytic expressions for the Lanczos
coefficients $ \alpha_n(N) $ and $ \beta^2_n(N) $ can be found at an
arbitrary Lanczos step $ n $ in terms of low order cumulants,
denoted by $ c_n \ n=1,\ldots $.
This expansion, also called the ``plaquette expansion''\cite{h93a},
\cite{wh94}, is actually an expansion of the exact coefficients about
the infinite size limit, that is in $ 1/N $.
Furthermore in the combined limit of complete Lanczos convergence
$ n \to \infty $ and the thermodynamic limit $ N \to \infty $
scaled Lanczos coefficients emerge in terms of $ z=n/N $, taken to
the lowest three orders
\begin{eqnarray}
   \alpha(z) 
 &\equiv&
    \lim_{n,N\rightarrow\infty} \alpha_n(N)
 \nonumber \\
 &=&
   c_1 + z\,\left[{c_{3}\over c_{2}}\right]
    + z^2\,\left[{3 c_3^3 - 4 c_2 c_3 c_4
                  + c_2^2 c_5 \over 4 c_2^4}\right] + O(z^3),
 \nonumber\\
    \beta^2(z) \vphantom{\Biggl[}
 &\equiv&
    \lim_{n,N\rightarrow\infty} \beta^2_n(N)
 \nonumber \\
 &=&
    z\,c_{2} 
    + z^2\left[{c_{2}c_{4} - c_{3}^{2}\over 2 c_{2}^{2}}\right]
\nonumber\\
 & &
    + z^3 \left[{21 c_{2} c_{3}^2 c_{4}-12 c_{3}^4 
                - 4 c_{2}^2 c_{4}^2 - 6 c_{2}^2 c_{3} c_{5} 
                    + c_{2}^3 c_{6} \over 12 c_{2}^5}\right] + O(z^4). 
\end{eqnarray}
In Ref~\cite{hw96} it was found that the ground-state
energy density in the bulk limit to be given in general by 
\begin{equation}
\epsilon_0 =  \inf_z\,\left[\alpha(z) - 2\,\beta(z)\right]
               \equiv  \inf_z\,e(z) \ . 
\end{equation}

All these moment methods use the same input data, a sequence of
connected moments up to some cut-off order, yet have rather different
treatments of these so that it is of interest to compare them with
each other.
This work is the logical extension of Ref\cite{woh95} which compared
the {\it t}-expansion and the connected moments expansion with
the coupled-cluster method, series expansions and spin-wave theory
for the same model.
In this work we have used the same set of cumulants as in the
above work, up to the 15th order, and an additional set taken to the
same order.
However the interest of this work is not just confined to a discussion
of the relative merits of the various methods, but also to demonstrate
the contribution of moment methods to an understanding of the planar
antiferromagnetic Heisenberg model.

\section{Ground State Averages for analytic Lanczos Expansion}
Most previous applications of the plaquette expansion have been to
the energy spectrum, 
notably the ground state energy\cite{h93b},\cite{th94},\cite{wh96} 
and the mass gap\cite{hww95}.
Two earlier examples of the application of the expansion to other
averages were the calculation of the staggered magnetisation of the
isotropic 2D Heisenberg model\cite{b93},\cite{ht94}. In these works
the staggered magnetisation was extracted from the full nonlinear
magnetic field dependence of the ground state energy for a system in
a transverse external field.
Here we take this approach to its logical conclusion and establish the
formalism to describe an 
ground state average of an arbitrary operator $ \hat{V} $.
Our approach is quite simple, we ``piggy-back'' our operator
$ \hat{V} $ onto our Hamiltonian, by tagging it with a variable
$ \lambda $, and apply the earlier results and theorems to this new 
Hamiltonian. 

Consider the exact ground state energy of the new Hamiltonian
$ \hat{H}+\lambda \hat{V} $, namely $ \epsilon(\lambda) $ and with
exact ground state wavefunction $ | \Psi_{\lambda}\rangle $.
\begin{equation}
   \epsilon(\lambda) = 
   \langle\Psi_{\lambda}| \hat{H}
                        +\lambda \hat{V} |\Psi_{\lambda}\rangle \ ,
\end{equation}
then the Hellmann-Feynman theorem gives the ground state average
as
\begin{eqnarray}
   \langle \hat{V}\rangle = \left. {d\over d\lambda}\epsilon(\lambda)
                      \right|_{\lambda=0} \ .
\end{eqnarray}
So we need only evaluate the first order change in $ \lambda $
in any quantity.
Because we are assuming all quantities thus derived are analytic in
some small neighbourhood of $ \lambda=0 $ we require the overlap
of the new ground state wavefunction with the original is nonzero,
$ \langle\Psi_0 | \Psi_{\lambda}\rangle \neq 0 $.

The new Hamiltonian $ H_{\lambda} $ will generate a new sequence
of Lanczos coefficients
$ \alpha_n(\lambda) $ and $ \beta^2_n(\lambda) $ from some suitable
trial state $ |\psi_0 \rangle $ and these in term will have analytic
expansions
\begin{eqnarray}
  \alpha_n(\lambda) 
 &=&
   \alpha_n + \lambda\delta^V\alpha_n + {\rm O}(\lambda^2)
 \nonumber \\
  \beta^2_n(\lambda) 
 &=&
   \beta^2_n + \lambda\delta^V\beta^2_n + {\rm O}(\lambda^2) \ ,
 \nonumber
\end{eqnarray}
where $ \delta^V\alpha_n $ and $ \delta^V\beta^2_n $ define the first 
order shifts in $ \alpha_n $ and $ \beta^2_n $ in the presence of
the operator $ \hat{V} $.
In the interests of clarity a symbol without an argument of
$ \lambda $ is assumed to be the case where $ \lambda=0 $.
Furthermore we consider models where the extensive scaling property
when the Lanczos iteration number $ n $ and the system size $ N $ both
tend to infinity with $ z=n/N $ fixed applies to the Lanczos
coefficients,
\begin{eqnarray}
  \alpha_n(\lambda) 
 &\stackrel{n,N \to \infty}{\to}&
   \alpha(z) + \lambda\delta^V\alpha(z) + {\rm O}(\lambda^2)
 \nonumber \\
  \beta^2_n(\lambda) 
 &\stackrel{n,N \to \infty}{\to}&
   \beta^2(z) + \lambda\delta^V\beta^2(z) + {\rm O}(\lambda^2) \ .
 \nonumber
\end{eqnarray}
Then the exact ground state theorem\cite{hw96} can be applied in the
form
\begin{equation}
  \epsilon(\lambda) = 
  \inf_{z}\left\{ \alpha(\lambda,z) 
                   - 2 [\beta^2(\lambda,z)]^{1/2} \right\} \ ,
\end{equation}
and one can show that
\begin{equation}
  \langle \hat{V}\rangle = \left. v(z) \right|_{z=\bar{z}} \ ,
\end{equation}
where our new $ z $-function $ v(z) $ is
\begin{equation}
  v(z) = \delta^V\alpha(z) - {\delta^V\beta^2(z) \over \beta(z)} \ ,
\end{equation}
and $ \bar{z} $ is the $ z $-value of the minima in 
$ e(z) = \alpha(z)-2\beta(z) $ if it exists.
The new quantities in this, our central result, are the first order
shifts in the Lanczos coefficients, and these will now be found in
terms of moments.

As is fundamental in the Lanczos process we need to find the moments
of the new Hamiltonian with respect to our trial state, and the first
order shifts of these $ T_n $, by
\begin{equation}
   \langle (\hat{H}_{\lambda})^n \rangle = 
   \langle \hat{H}^n \rangle + \lambda T_n + {\rm O}(\lambda^2) \ ,
\end{equation}
and it is found that $ T_n $ is given by a sum of distributed
generalised moments,
\begin{equation}
   T_n = \sum^{n-1}_{k=0}
         \langle \hat{H}^{n-1-k} \hat{V} \hat{H}^k \rangle \ ,
 \label{moment}
\end{equation}
when $ n>1 $. The first member, $ T_0=0 $.
This result can be proven from the recurrence relation 
$ T_{n+1} = \langle \hat{H}\hat{T_n}\rangle 
          + \langle \hat{V}\hat{H}^n \rangle $ 
or from the generating function
\begin{equation}
  \sum^{\infty}_{n=0} w^n T_n =
  w\langle {1\over 1-w\hat{H}} \hat{V} {1\over 1-w\hat{H}} \rangle \ .
\end{equation}
From this point we go to the connected parts of these moments, with a
first order shift denoted by $ S_n $,
\begin{equation}
   \langle (H_{\lambda})^n \rangle_c = 
   \langle \hat{H}^n \rangle_c + \lambda S_n + {\rm O}(\lambda^2) \ .
\end{equation}
This connected part, or cumulant, has the standard generating function
\begin{equation}
   \left.
   {d\over d\lambda} \log 
            \langle e^{t H_{\lambda}} \rangle\right|_{\lambda=0} =
                     \sum^{\infty}_{n=1} {t^n \over n!} S_n  \ ,
\end{equation}
and the standard recurrence relation with the moments
\begin{equation}
   T_{n+1} = \sum^n_{k=0} {n+1 \choose k}
                          \langle \hat{H}^k \rangle S_{n+1-k} \ ,
\end{equation}
for $ n \ge 0 $.
In effect what this reduces $ S_n $ to is a sum similar to 
Eq.~\ref{moment} except now it is 
over the connected parts of the distributed generalised moments,
namely $ \langle \hat{H}^{n-k} \hat{V} \hat{H}^k \rangle_c $. 
It is these moments which are actually directly calculated.
Finally, in the extensive many-body problem the connected moments have
a size dependence via the coefficients
\begin{equation}
  S_n = \delta c_n N \ .
\end{equation}

The expression for $ T_n $ (see Eq.~\ref{moment}) or $ S_n $
in terms of the distributed generalised 
moments bears some similarity with the ones appearing in the
{\it t}-expansion (see Equation 2.10 in Ref\cite{woh95}), in that
exactly the same averages occur, but that here the binomial
coefficient is absent.
This is because the Lanczos process is one where the Hamiltonian
$ \hat{H} $ generates new states in an enlarging state space, akin to
a geometrical progression, and geometrical generating functions and
convolutions arise.
In contrast the {\it t}-expansion employs an exponentially mapped
state $ e^{-\frac{1}{2}tH}|\psi_0 \rangle $ and therefore exponential
generating functions and convolutions appear.

It should be noted that all the forgoing is exact, given that a system
is solvable in this sense and all the moments,
and Lanczos coefficients are analytically known. 
However for many non-trivial models one has knowledge
only of the first say, $ 2r $ moments which give terms of  
$ z^0, \ldots, z^{r-1} $ in $ \alpha(z) $ and $ z, \ldots, z^r $ in 
$ \beta^2(z) $, or $ 2r+1 $ moments which give terms in $ \alpha(z) $
from $ z^0, \ldots, z^r $ and terms in $ \beta^2(z) $ from
$ z, \ldots, z^r $.
In the even case, with $ 2r $ moments, we define the truncation order
as $ r $ and this is a natural ordering as there are equal numbers of
terms in $ \alpha $ and $ \beta^2 $ coefficients, 
and in the odd case $ 2r+1 $ the truncation order is still $ r $, 
but is not a natural order, but termed supplemented, in having one
more term, namely a $ z^r $ in $ \alpha(z) $.

Given the shifted cumulant coefficients $ \delta c_n $, and therefore
the shifted moments $ T_n $ there exists a number of algorithms for
generating the shifted Lanczos coefficients, and one of the more
robust and efficient ones is briefly described in the Appendix.
The first few terms in the Lanczos coefficients are given by
\begin{eqnarray}
 \delta^V\alpha(z)
 &=&
   \delta c_1 
   + z \left( -\delta c_2{c_3 \over c^2_2}+\delta c_3{1\over c_2}
       \right)
 \nonumber\\
 && \quad
   + \frac{1}{2} z^2 \left(
     \delta c_2 \left[ -6{c^3_3 \over c^5_2}+6{c_3c_4 \over c^4_2}
                       -{c_5 \over c^2_2} \right]
    +\delta c_3 \left[ \frac{9}{2}{c^2_3 \over c^4_2}
                                -2{c_4 \over c^3_2} \right]
    +\delta c_4 \left[ -2{c_3 \over c^3_2} \right]
    +\delta c_5 \left[ {1 \over 2c^2_2} \right]
                     \right)
 \nonumber\\
 && \quad \quad
   + {\rm O}(z^3) \ ,
 \nonumber\\
 \delta^V\beta^2(z)
 &=&
   z \delta c_2 
 \nonumber \\
 && \quad
   + \frac{1}{2} z^2 \left(
     \delta c_2 \left[ 2{c^2_3 \over c^3_2}-{c_4 \over c^2_2} \right]
    +\delta c_3 \left[ -2{c_3 \over c^2_2} \right]
    +\delta c_4 \left[ {1 \over c_2} \right]
                     \right)
 \nonumber\\
 && \quad \quad
   + \frac{1}{6} z^3 \left(
     \delta c_2 \left[ 30{c^4_3 \over c^6_2}-42{c^2_3c_4 \over c^5_2}
                       +6{c^2_4 \over c^4_2}+9{c_3c_5 \over c^4_2}
                       -{c_6 \over c^3_2} \right]
    +\delta c_3 \left[ -24{c^3_3 \over c^5_2}+21{c_3c_4 \over c^4_2}
                       -3{c_5 \over c^3_2} \right]
                     \right.
 \nonumber\\
 && \quad \quad
                     \left.
   \phantom{+ \frac{1}{6} z^3 \left(\right.}
    +\delta c_4 \left[ \frac{21}{2}{c^2_3 \over c^4_2}
                                 -4{c_4 \over c^3_2} \right]
    +\delta c_5 \left[ -3{c_3 \over c^3_2} \right]
    +\delta c_6 \left[ {1 \over 2c^2_2} \right]
                     \right)
 \nonumber\\
 && \quad \quad \quad
   + {\rm O}(z^4) \ .
\end{eqnarray}

Lastly we report some similar results to the above for the excited
state gap, which was first found at the level of the first order
plaquette expansion in Ref\cite{hww95}, and has been generalised in
Ref.\cite{wmh96}.
The connected moments for the excited (triplet) state $ c^P_n N $
are related to the ground (singlet) state cumulants $ c^S_n N $ by
\begin{equation}
        c^P_n N = c^S_n N + \delta^G c_n \ .
\end{equation}
One can define shifted Lanczos coefficients by
\begin{eqnarray}
 \delta^G\alpha(z)
 &=&
 \lim_{n,N \to \infty} N \left\{ \alpha^P_n(N)-\alpha^S_n(N) \right\}
 \nonumber\\
 \delta^G\beta^2(z)
 &=&
 \lim_{n,N \to \infty} 
         N \left\{ [\beta^P_n(N)]]^2-[\beta^S_n(N)]^2 \right\} \ ,
\end{eqnarray}
and there is an analogous exact result for the triplet gap
\begin{equation}
   g = \left. m(z) \right|_{z=\bar{z}} \ ,
\end{equation}
where the gap function $ m(z) $ is defined as
\begin{equation}
  m(z) = \delta^G\alpha(z) - {\delta^G\beta^2(z) \over \beta(z)} \ .
\end{equation}
The shifted cumulants $ \delta^G c_n $ can be found from the
equivalent {\it t}-expansion function $ R(t) $\cite{woh95} via
\begin{equation}
   \log R(t) \equiv \log{ \langle\psi^P_0 | e^{-tH} |\psi^P_0 \rangle
                          \over
                          \langle\psi^S_0 | e^{-tH} |\psi^S_0 \rangle
                        }
             = \sum^{\infty}_{n=1} {(-t)^n \over n!} \delta c_n \ .
\end{equation}

\section{Application to the 2D XXZ Heisenberg Model}
We apply the plaquette expansion formalism to the $ s=1/2 $
anisotropic antiferromagnetic Heisenberg Model on a square lattice,
with the usual Hamiltonian
\begin{equation}
H=\sum_{<i,j>} \left[ S^z_i S^z_j
             + x\Bigl( S^x_i S^x_j + S^y_i S^y_j \Bigr) \right] \ ,
\end{equation}
where $ x=0 $ denotes the Ising Model and $ x=1 $ the isotropic
antiferromagnetic Heisenberg Model.
For the trial ground state we take the ``natural'' and commonly
employed choice of the eigenstate to $ x=0 $ problem,
the classical Neel state, and for the excited triplet trial state a
state with one spin of the Neel state flipped.
We have used the cumulants $ I_n $ and $ R_n $ in Ref.\cite{woh95},
and generated a new set of staggered magnetisation cumulants to an
order equivalent to that used in the above reference.
From these we have generated the Lanczos coefficients
and the staggered magnetisation Lanczos coefficients up to 7th order,
and the triplet gap Lanczos coefficients up to 5th order.
All new cumulants and the Lanczos coefficients are displayed in the
Appendix.

Using these cumulants the ground state energy density $ \epsilon_0 $,
the staggered magnetisation $ \cal M $ and
triplet gap $ \cal G $ have been computed for a set of
anisotropy parameters and for all truncation orders up to the maximum,
by evaluation at the minima of $ e(z) $.
This is the simplest way of analysing the plaquette expansion and
is free from any biasing and assumptions, so it is the preferred
option. Given more understanding of the convergence properties of
the plaquette expansion a more sophisticated analysis or extrapolation
strategy may be employed, and we refer the reader to Ref\cite{wh96}
for a discussion of this issue.
The results for the ground state energy density, the staggered
magnetisation and the excited state gap are displayed
in Table~\ref{gse-table}, Table~\ref{mag-table}, and 
Table~\ref{gap-table} respectively.
Depending on the order and anisotropy parameter, the minima in the
ground state energy function $ e(z) $, could be complex, usually close
to the positive real axis, however. 
In these cases we took the real part of the
function computed at the complex minima, and these are marked with a
asterix(*).
To treat the case of no real minima properly some form of extrapolation
would be needed but we do not pursue this issue here for several
reasons. Firstly, a real minima exists for all the data at the 
highest order ($ r=7 $), which is the most important, at least for 
the ground state energy and magnetisation.
Secondly there is some continuity at a given order, between the
data where a real minima exists and that data where it is complex.
As the anisotropy parameter varies through these regions the
minima shifts off slightly from the real axis but nonetheless
remains close.
There is also a second number appended to each data entry of the 
tables, and that is the difference between that value computed 
using the natural ordering and the supplemented ordering.
This is the most unbiased estimate we have for indicating the
systematic error in our expansion although there are no rigorous
results to say that this difference somehow bounds the error - 
it is at best semi-quantitative.

The plaquette expansion results display two trends - one where as the
truncation order increases we find substantial and systematic
improvement in the averages, at least over the region where a real
minima exists, and another as the anisotropy parameter increases from
zero, where we find rapidly increasing errors, although the extent of
this is dependent on the particular quantity.
The ground state energy density is the most accurate, then the
staggered magnetisation, and the triplet gap is the worst.

\section{Comparison of the Methods}
Taking the highest order plaquette expansion results, we have
tabulated these against the other methods in 
Table~\ref{compare-gse-table}, Table~\ref{compare-mag-table} and 
Table~\ref{compare-gap-table}.

For the ground state energy one can see clear trends in the
comparison :
\begin{itemize}
 \item
 for $ x < 0.5 $ the {\it t}-expansion and CMX are slightly better
 than the plaquette expansion although the actual differences between
 them and with the series results are very small,
 \item
 for $ x = 0.8 $ the plaquette expansion is better than the
 {\it t}-expansion and the CMX, while for $ x = 0.9 $ the
 {\it t}-expansion is a slight improvement over the plaquette
 expansion,
 \item
 and for $ x > 0.95 $ the plaquette expansion is clearly better than
 the {\it t}-expansion, the CMX and 3rd order spin-wave theory, and
 especially so at $ x = 1 $.
\end{itemize}

For the staggered magnetisation one observes the following trends -
\begin{itemize}
 \item
 for $ x = 0.5 $ the plaquette expansion is better than the
 {\it t}-expansion and the CMX,
 \item
 for $ x = 0.8 $ this ordering has reversed with the {\it t}-expansion
 superior to the plaquette expansion and the CMX, while for
 $ x = 0.9 $ the order gets reversed once more, and the plaquette
 expansion is a slight improvement over the {\it t}-expansion,
 \item
 and for $ x \to 1.0 $ the plaquette expansion stabilises on a higher
 value than that given by the {\it t}-expansion, 3rd order spin-wave
 theory, and the series expansion, although remaining less than the
 CMX result.
\end{itemize}

And for the triplet gap one finds that all moment methods give a
rather high nonzero value at the isotropic point, 
whereas its vanishing for the spin-wave theory and series expansion is
built-in. 
We would like to emphasise that for $ x \ge 0.9 $ there was no real
minima at the highest order that applied in the plaquette expansion, 
and consequently the estimated errors grossly underestimate the true
error. The trends here are -
\begin{itemize}
 \item
 for $ x = 0.2 \to 0.5 $ the {\it t}-expansion is better than the CMX
 and the plaquette expansion,
 \item
 for $ x = 0.8 \to 0.9 $ the {\it t}-expansion is still superior to
 the plaquette expansion although it, in turn is better than the CMX,
 \item
 and for $ x > 0.95 $ the plaquette expansion is slightly better than
 the other moment methods but all have errors which swamp the actual
 values.
\end{itemize}

Finally we compare these results with the coupled-cluster results
at the isotropic point. The results given in 
Table~\ref{compare-gse-table} and Table~\ref{compare-mag-table}
are the extrapolated results based on the first few orders of the
approximation, and differ significantly from the highest order
results. Both recent extrapolated ground state energies are better
than either the $t$-expansion or the CMX and comparable in accuracy
to the plaquette expansion energy. The extrapolated staggered
magnetisations are also very close to that predicted by the 
plaquette expansion, probably higher than the true value.

The relatively poor performance of the {\it t}-expansion arises from a
number of reasons. Firstly the extrapolation $ t \to \infty $ is done
without any knowledge of the global analytic properties of $ E(t) $,
and it is widely known in extrapolation work that this is perhaps the
largest single source of poor convergence and inaccuracies.
In contrast there is no extrapolation problem in the plaquette
expansion, so long as a real minima exists. If this is not the case
then some form of extrapolation of the kind employed in 
Ref\cite{wh96} may be necessary here.

Another source of mediocre performance, and this is shared with the
plaquette expansion, is the rather poor quality of trial state.
The classical Neel state is usually chosen because it is simple and
relatively straightforward to generate moments with, but a trial state
which can generate the correct singularity structure at the isotropic
point may give better results even though fewer moments may be found
from it.

In summary we found good quantitative predictions from the plaquette
expansion for this model but, as expected, the worst results occured
near the isotropy point where a first order transition to a gapless
state is expected.
The plaquette expansion was clearly better at predicting the ground
state energy than the other moment methods, had a slighter higher
staggered magnetisation than the others, and all of the methods gave a
nonzero gap at the isotropic point.
We would like to emphasise that many aspects of the moment methods are
still poorly understood, and that considerable scope exists for
improving their accuracy.

We would like to acknowledge that this work was supported by the
Australian Research Council.

%
% is shown in Fig.~\ref{typical-case-AH} 
% and Fig.~\ref{typical-case-XY} respectively
%
\newpage
\section{Appendix}
Using efficient graph enumeration and calculation of the embedding
constants Ref.~\cite{woh95},\cite{hho90}
we have the shift in the staggered magnetisation cumulants
\begin{eqnarray}
 \delta^M c_1
 &=&
    1/2
 \nonumber \\
 \delta^M c_2
 &=&
    0
 \nonumber \\
 \delta^M c_3
 &=&
    -x^{2}
 \nonumber \\
 \delta^M c_4
 &=&
    -6\,x^{2}
 \nonumber \\
 \delta^M c_5
 &=&
    10\,x^{4}-27\,x^{2}
 \nonumber \\
 \delta^M c_6
 &=&
    231\,x^{4}-108\,x^{2}
 \nonumber \\
 \delta^M c_7
 &=&
    -266\,x^{6}+3153\,x^{4}-405\,x^{2}
 \nonumber \\
 \delta^M c_8
 &=&
    -13382\,x^{6}+33494\,x^{4}-1458\,x^{2}
 \nonumber \\
 \delta^M c_9
 &=&
    12953\,x^{8}-374132\,x^{6}+307212\,x^{4}-5103\,x^{2}
 \nonumber \\
 \delta^M c_{10}
 &=&
    1153345\,x^{8}-7722995\,x^{6}+2562205\,x^{4}-17496\,x^{2}
 \nonumber \\
 \delta^M c_{11}
 &=&
    -{\frac {3989639\,x^{10}}{4}}+{\frac {109952063\,x^{8}}{2}}
            -131870963\,x^{6}+20015055\,x^{4}-59049\,x^{2}
 \nonumber \\
 \delta^M c_{12}
 &=&
    -{\frac {559451229\,x^{10}}{4}}+{\frac {3755433743\,x^{8}}{2}}
             -1977566959\,x^{6}+149166828\,x^{4}-196830\,x^{2}
 \nonumber \\
 \delta^M c_{13}
 &=&
    {\frac {446772107\,x^{12}}{4}}-{\frac {20445508149\,x^{10}}{2}}
            +{\frac {103437271579\,x^{8}}{2}}
 \nonumber \\
 & & \qquad
    -27001597819\,x^{6}+1073575434\,x^{4}-649539\,x^{2}
 \nonumber \\
 \delta^M c_{14}
 &=&
    {\frac {182477940639\,x^{12}}{8}}
          -{\frac {2097642068485\,x^{10}}{4}}
                 +{\frac {2447876418449\,x^{8}}{2}}
 \nonumber \\
 & & \qquad
    -343658374738\,x^{6}+7524115139\,x^{4}-2125764\,x^{2}
 \nonumber \\
 \delta^M c_{15}
 &=&
    -{\frac {68595645155\,x^{14}}{4}}
           +{\frac {4764540794257\,x^{12}}{2}}
                  -{\frac {170554817150787\,x^{10}}{8}}
 \nonumber \\
 & & \qquad
    +25875363869594\,x^{8}-4143420881204\,x^{6}
                          +51653871357\,x^{4}-6908733\,x^{2}
\end{eqnarray}
and using the excited state coefficients from Ref.\cite{woh95}
we have the shift in the excited state cumulants
\begin{eqnarray}
 \delta^G c_{1}
 &=&
    2
 \nonumber \\
 \delta^G c_{2}
 &=&
    2\,x^{2}
 \nonumber \\
 \delta^G c_{3}
 &=&
    0
 \nonumber \\
 \delta^G c_{4}
 &=&
    -8\,x^{4}-12\,x^{2}
 \nonumber \\
 \delta^G c_{5}
 &=&
    -46\,x^{4}-60\,x^{2}
 \nonumber \\
 \delta^G c_{6}
 &=&
    209\,x^{6}+115\,x^{4}-228\,x^{2}
 \nonumber \\
 \delta^G c_{7}
 &=&
    3962\,x^{6}+4798\,x^{4}-780\,x^{2}
 \nonumber \\
 \delta^G c_{8}
 &=&
    -{\frac {40123\,x^{8}}{4}}+{\frac {132689\,x^{6}}{4}}+59703\,x^{4}
             -2532\,x^{2}
 \nonumber \\
 \delta^G c_{9}
 &=&
    -{\frac {795907\,x^{8}}{2}}-{\frac {204405\,x^{6}}{2}}
             +558754\,x^{4}-7980\,x^{2}
 \nonumber \\
 \delta^G c_{10}
 &=&
    {\frac {3252469\,x^{10}}{4}}-{\frac {33276617\,x^{8}}{4}}
            -9011003\,x^{6}+4548463\,x^{4}-24708\,x^{2}
 \nonumber \\
 \delta^G c_{11}
 &=&
    {\frac {109203791\,x^{10}}{2}}-{\frac {193681633\,x^{8}}{2}}
            -{\frac {394547565\,x^{6}}{2}}
 \nonumber \\
 & & \quad
    +34055630\,x^{4}-75660\,x^{2}
 \nonumber \\
 \delta^G c_{12}
 &=&
    -88838574\,x^{12}+754813869\,x^{10}
                     +5952720589\,x^{8}+3261131208\,x^{6}
 \nonumber \\
 & & \quad
    -4001975480\,x^{4}+83156352\,x^{2}-4096
\end{eqnarray}

The expansion for the Lanczos coefficients are given exactly
\cite{ggps83},\cite{ggp85} up to their truncation order by
\begin{eqnarray}
 \alpha(z) =
 &&
-1/2
 \nonumber \\
 &&
+3\,z
 \nonumber \\
 &&
-{\frac {11\,z^{2}}{2}}
 \nonumber \\
 &&
 +\left (
        -{\frac {89}{18}}
        -{\frac {68}{9\,x^{2}}}
 \right )z^{3}
 \nonumber \\
 &&
+\left (
        -{\frac {167}{72}}
        +{\frac {397}{18\,x^{2}}}
        -{\frac {298}{9\,x^{4}}}
 \right )z^{4}
 \nonumber \\
 &&
+\left (
        +{\frac {45967}{1200}}
        +{\frac {14257}{180\,x^{2}}}
        +{\frac {58376}{225\,x^{4}}}
        -{\frac {9784}{45\,x^{6}}}
 \right )z^{5}
 \nonumber \\
 &&
+\left (
        +{\frac {2668781}{10800}}
        +{\frac {19041737}{21600\,x^{2}}}
        +{\frac {704387}{1800\,x^{4}}}
        +{\frac {5758538}{2025\,x^{6}}}
        -{\frac {144088}{81\,x^{8}}}
 \right )z^{6}
 \nonumber \\
 &&
+\left (
        +{\frac {40546957}{52920}}
        +{\frac {10673263429}{12700800\,x^{2}}}
        +{\frac {2863042529}{352800\,x^{4}}}
% \right.
% \nonumber \\
% &&
% \hskip5.0cm
% \left.
        -{\frac {234213799}{99225\,x^{6}}}
        +{\frac {14449160}{441\,x^{8}}}
        -{\frac {1356032}{81\,x^{10}}}
 \right )z^{7}
\end{eqnarray}
and
\begin{eqnarray}
 \beta^2(z) =
 &&
+{\frac {x^{2}z}{2}}
 \nonumber \\
 &&
        -{\frac {5\,x^{2}z^{2}}{4}}
 \nonumber \\
 &&
+\left (
        +{\frac {5}{2}}
        +{\frac {7\,x^{2}}{6}}
 \right )z^{3}
 \nonumber \\
 &&
+\left (
        -{\frac {29}{6}}
        +{\frac {137\,x^{2}}{72}}
        +{\frac {26}{3\,x^{2}}}
 \right )z^{4}
 \nonumber \\
 &&
+\left (
        -{\frac {422}{45}}
        +{\frac {565\,x^{2}}{288}}
        -{\frac {1613}{30\,x^{2}}}
        +{\frac {254}{5\,x^{4}}}
 \right )z^{5}
 \nonumber \\
 &&
 +\left (
        -{\frac {2961379}{21600}}
        -{\frac {361171\,x^{2}}{57600}}
        -{\frac {160703}{2160\,x^{2}}}
        -{\frac {388447}{675\,x^{4}}}
        +{\frac {3496}{9\,x^{6}}}
 \right )z^{6}
 \nonumber \\
 &&
+\left (
        -{\frac {1465471213}{3628800}}
        -{\frac {18027853\,x^{2}}{362880}}
        -{\frac {3540413}{2100\,x^{2}}}
% \right.
% \nonumber \\
% &&
% \hskip5.0cm
% \left.
        +{\frac {31611547}{113400\,x^{4}}}
        -{\frac {91790458}{14175\,x^{6}}}
        +{\frac {94360}{27\,x^{8}}}
 \right )z^{7}
\end{eqnarray}

The expansion for the change in the Lanczos coefficients corresponding
to the staggered magnetisation is given by
\begin{eqnarray}
 \delta^{SM}\alpha(z) =
 &&
         1/2
 \nonumber \\
 &&
        -2\,z
 \nonumber \\
 &&
+\left (
        +{\frac {19}{9}}
        +{\frac {10}{3\,x^{2}}}
 \right )z^{3}
 \nonumber \\
 &&
+\left (
        +{\frac {23}{2}}
        +{\frac {376}{9\,x^{2}}}
        +{\frac {46}{x^{4}}}
 \right )z^{4}
 \nonumber \\
 &&
+\left (
        +{\frac {32653}{900}}
        +{\frac {43007}{900\,x^{2}}}
        +{\frac {20827}{75\,x^{4}}}
        +{\frac {40064}{75\,x^{6}}}
 \right )z^{5}
 \nonumber \\
 &&
+\left (
        +{\frac {1852433}{64800}}
        -{\frac {37342427}{32400\,x^{2}}}
        -{\frac {6621893}{2700\,x^{4}}}
        +{\frac {991187}{2025\,x^{6}}}
        +{\frac {286984}{45\,x^{8}}}
 \right )z^{6}
 \nonumber \\
 &&
+\left (
        -{\frac {55919677}{127008}}
        -{\frac {21521839999}{2116800\,x^{2}}}
        -{\frac {2231398105}{63504\,x^{4}}}
% \right.
% \nonumber \\
% &&
% \hskip4.5cm
% \left.
        -{\frac {10777758037}{198450\,x^{6}}}
        -{\frac {899068624}{33075\,x^{8}}}
        +{\frac {75159136}{945\,x^{10}}}
 \right )z^{7}
 \nonumber \\
 &&
\end{eqnarray}
and
\begin{eqnarray}
 \delta^{SM}\beta^2(z) =
 &&
 \left (
        -{\frac {64}{9}}
        -{\frac {68}{9\,x^{2}}}
 \right )z^{4}
 \nonumber \\
 &&
+\left (
        -{\frac {871}{45}}
        -{\frac {3959}{60\,x^{2}}}
        -{\frac {290}{3\,x^{4}}}
 \right )z^{5}
 \nonumber \\
 &&
+\left (
        +{\frac {442903}{10800}}
        +{\frac {328027}{1350\,x^{2}}}
        -{\frac {2500}{9\,x^{4}}}
        -{\frac {10520}{9\,x^{6}}}
 \right )z^{6}
 \nonumber \\
 &&
+\left (
        +{\frac {734390231}{907200}}
        +{\frac {526210073}{113400\,x^{2}}}
        +{\frac {117862772}{14175\,x^{4}}}
        +{\frac {40289357}{14175\,x^{6}}}
        -{\frac {5888216}{405\,x^{8}}}
 \right )z^{7}
\end{eqnarray}

The expansion for the shift in the Lanczos coefficients describing
the triplet energy gap is given by
\begin{eqnarray}
 \delta^{G}\alpha(z) =
 &&
        +2
 \nonumber \\
 &&
        -12\,z
 \nonumber \\
 &&
+\left (
        +34
        -{\frac {24}{x^{2}}}
 \right )z^{2}
 \nonumber \\
 &&
+\left (
        -{\frac {388}{3}}
        +{\frac {4118}{9\,x^{2}}}
        -{\frac {448}{3\,x^{4}}}
 \right )z^{3}
 \nonumber \\
 &&
+\left (
        +{\frac {52681}{72}}
        -{\frac {77915}{24\,x^{2}}}
        +{\frac {44726}{9\,x^{4}}}
        -{\frac {3920}{3\,x^{6}}}
 \right )z^{4}
\end{eqnarray}
and
\begin{eqnarray}
 \delta^{G}\beta^2(z) =
 &&
        +2\,x^{2}z
 \nonumber \\
 &&
+\left (
        -3\,x^{2}
        +6
 \right )z^{2}
 \nonumber \\
 &&
+\left (
        +{\frac {71\,x^{2}}{3}}
        -{\frac {239}{3}}
        +{\frac {32}{x^{2}}}
 \right )z^{3}
 \nonumber \\
 &&
+\left (
        -{\frac {1943\,x^{2}}{72}}
        +{\frac {3929}{8}}
        -{\frac {5413}{6\,x^{2}}}
        +{\frac {784}{3\,x^{4}}}
 \right )z^{4}
 \nonumber \\
 &&
+\left (
        +{\frac {143947\,x^{2}}{720}}
        -{\frac {164471}{48}}
        +{\frac {710417}{72\,x^{2}}}
        -{\frac {481214}{45\,x^{4}}}
        +{\frac {7840}{3\,x^{6}}}
\right )z^{5}
\end{eqnarray}
\newpage
\begin{table}
\caption{
The ground state energy $ \epsilon_0 $ and estimated error for
the 2D XXZ Model calculated using the plaquette expansion
as a function of the anisotropy parameter $x$ and the order of
truncation $r$. Those energies where one has the case
of a real minima are unmarked, whereas those energies which are
given by the real part evaluated at a complex minima are marked
by a asterix. }
\label{gse-table}
\bigskip
\begin{tabular}{ccccccc}
 $\epsilon_0$ & $r=2$ & 3     & 4     & 5     & 6     & 7     \\
\hline
 $x=0.2$
& -0.50663(3)
& -0.506661(2)
& -0.5066639(5)
& -0.5066646(2)
& -0.50666494(8)
& -0.50666508(4) \\

    0.5
& -0.540(1)
& -0.5415(1)
& -0.54159(2)
& -0.541617(6)
& -0.541627(3)
& -0.541631(1) \\

    0.8
& -0.599(5)
& -0.605(1)
& -0.6068(2)
& -0.606975(6)
& -0.60697(2)
& -0.60693(2) \\

    0.9
& -0.622(8)
& -0.633(2)
& -0.6363(3) *
& -0.63667(7) *
& -0.6365(2) *
& -0.6360(1) \\

    0.95
& -0.64(1)
& -0.648(4)
& -0.6526(2) *
& -0.65280(3) *
& -0.653(1) *
& -0.6521(4) \\

    0.98
& -0.64(1)
& -0.658(4)
& -0.6627(1) *
& -0.66276(2) *
& -0.663(1) *
& -0.6622(5) \\

    0.99
& -0.65(1)
& -0.661(5)
& -0.66604(9) *
& -0.66613(4) *
& -0.666(1) *
& -0.6657(6) \\

    1.0
& -0.65(1)
& -0.664(5)
& -0.66945(7) *
& -0.66952(7) *
& -0.670(1) *
& -0.6691(6) \\

\end{tabular}
\end{table}
\newpage
\begin{table}
\caption{
The staggered magnetisation $ \cal M $ and estimated error for
the 2D XXZ Model as calculated using the plaquette expansion as
a function of the anisotropy parameter $x$ and the order of
truncation $r$. Again those energies where one has the case
of a real minima are unmarked, whereas those energies which are
given by the real part evaluated at a complex minima are marked
by a asterix. }
\label{mag-table}
\bigskip
\begin{tabular}{ccccccc}
 $\cal M $  & $r=2$ & 3     & 4     & 5     & 6     & 7     \\
\hline
 $x=0.2$
&  0.49563(7)
&  0.495543(8)
&  0.495532(2)
&  0.4955293(8)
&  0.4955281(4)
&  0.4955275(2) \\

    0.5
&  0.475(2)
&  0.4717(4)
&  0.47112(8)
&  0.47100(2)
&  0.470960(9)
&  0.470946(4) \\

    0.8
&  0.44(1)
&  0.424(6)
&  0.416(1)
&  0.4143(4)
&  0.4152(6)
&  0.4162(4) \\

    0.9
&  0.43(2)
&  0.40(2)
&  0.373(5) *
&  0.3794(4) *
&  0.377(3) *
&  0.383(3) \\

    0.95
&  0.43(2)
&  0.39(3)
&  0.367(4) *
&  0.371(2) *
&  0.37(1) *
&  0.366(6) \\

    0.98
&  0.42(3)
&  0.38(2)
&  0.364(3) *
&  0.367(3) *
&  0.359(1) *
&  0.358(5) \\

    0.99
&  0.42(3)
&  0.37(2)
&  0.363(3) *
&  0.365(3) *
&  0.35676(7) *
&  0.355(5) \\

    1.0
&  0.42(3)
&  0.37(2)
&  0.362(3) *
&  0.364(4) *
&  0.355(4) *
&  0.353(5) \\

\end{tabular}
\end{table}
\newpage
\begin{table}
\caption{
The excited state gap $ \cal G $ and estimated error for the
2D XXZ Model calculated using the plaquette expansion as a
function of the anisotropy parameter $x$ and the order of
truncation $r$. Again those energies where one has the case
of a real minima are unmarked, whereas those energies which are
given by the real part evaluated at a complex minima are marked
by a asterix. }
\label{gap-table}
\bigskip
\begin{tabular}{ccccc}
 $\cal G $  & $r=2$ & 3     & 4     & 5     \\
\hline
 $x=0.2$
&  1.943(4)
&  1.938(1)
&  1.936(5)
&  1.936 \\

    0.5
&  1.67(4)
&  1.617(7)
&  1.607(2)
&  1.603 \\

    0.8
&  1.2(1)
&  1.03(5)
&  0.962(9)
&  0.953 \\

    0.9
&  1.1(2)
&  0.8(2)
&  0.56(5) *
&  0.61 *\\

    0.95
&  1.0(2)
&  0.6(2)
&  0.47(3) *
&  0.51 *\\

    0.98
&  1.0(2)
&  0.5(2)
&  0.42(3) *
&  0.45 *\\

    0.99
&  1.0(3)
&  0.5(1)
&  0.40(2) *
&  0.42 *\\

    1.0
&  1.0(3)
&  0.5(1)
&  0.39(2) *
&  0.40 *\\

\end{tabular}
\end{table}
\newpage
\begin{table}
\caption{
Comparison of the most accurate plaquette expansion values for
the ground state energy $ \epsilon_0 $ with those of the
{\it t}-expansion, the connected moments expansion, spin-wave
theory and series expansions, taken for various anisotropy
parameters $x$. }
\label{compare-gse-table}
\bigskip
\begin{tabular}{cccccccc}
 $\epsilon_0 $
              & plaquette
              & {\it t}-expansion\tablenote{Reference~\cite{woh95}}
              & {\it t}-expansion\tablenote{Reference~\cite{woh95}}
              & CMX\tablenote{Reference~\cite{woh95}}     
              & 3rd Order\tablenote{Reference~\cite{hza92}}
              & series\tablenote{Reference~\cite{zoh91}}
              & coupled\tablenote{Reference~\cite{h93},\cite{bhx94}} \\
              & expansion 
              & D Pad\'e      
              & Laplace
              & 
              & spin-wave
              & expansion
              & cluster  \\
\hline
$x=0.2$	
&  -0.5066651(1)	& -0.5066653 	& -0.5066653	
& -0.50666529		& -0.50657179	& -0.5066653
&	\\
   0.5 	
&  -0.541631(1)	& -0.541636(3) 	& -0.5416359	
& -0.54163641	& -0.5413803	& -0.5416371
&	\\
   0.8 	
&  -0.60693(2)	& -0.6068(2) 	& -0.6067604	
& -0.60677223	& -0.607376	& -0.606902(2)
&	\\
   0.9 	
&  -0.6360(2)	& -0.6357(2) 	& -0.6353801	
& -0.63537633	& -0.636654	& -0.635844(4)
&	\\
   0.95	
&  -0.6521(2)	& -0.6518(3) 	& -0.6510589	
& -0.65101764	& -0.652718	& -0.65189(1)
&	\\
   0.98	
&  -0.6622(2)	& -0.6618(4) 	& -0.6609227	
& -0.66083842	& -0.66287	& -0.66211(2)
&	\\
   0.99	
&  -0.6656(3)	& -0.6651(6) 	& -0.6642888	
& -0.66418527	& -0.66637	& -0.66563(6)
&	\\
   1.0 	
&  -0.6691(3)	& -0.668(1) 	& -0.6676946	
& -0.66756890	& -0.6699993	& -0.6693(1)
& -0.6692 	\\

&            	&           	&           	
&            	&           	&           
& -0.6691(3)	\\
\end{tabular}
\end{table}
\begin{table}
\caption{
Comparison of the most accurate plaquette expansion values for
the staggered magnetisation $ \cal M $ with those of the
{\it t}-expansion, the connected moments expansion, spin-wave
theory and series expansions, taken for various anisotropy
parameters $x$. }
\label{compare-mag-table}
\bigskip
\begin{tabular}{ccccccc}
 $\cal M $    & plaquette
              & {\it t}-expansion\tablenote{Reference~\cite{woh95}}
              & CMX\tablenote{Reference~\cite{woh95}}     
              & 3rd Order\tablenote{Reference~\cite{hza92}}
              & series\tablenote{Reference~\cite{zoh91}}
              & coupled\tablenote{Reference~\cite{h93},\cite{bhx94}} \\
              & expansion 
              & D Pad\'e      
              &
              & spin-wave
              & expansion
              & cluster  \\
\hline
$x=0.2$	
&  0.4955275(2)	& 0.495527(1) 	&  0.49552690	
& 0.49573699	& 0.4955265	&		\\
   0.5 	
&  0.470946(4)	& 0.4710(4) 	&  0.47097192	
& 0.47172243	& 0.4709287	&		\\
   0.8 	
&  0.4162(4)	& 0.4173(6) 	&  0.41727472	
& 0.416390	& 0.416896(5)	&		\\
   0.9 	
&  0.383(3)	& 0.386(4) 	&  0.39065099	
& 0.383864	& 0.38553(2)	&		\\
   0.95	
&  0.365(3)	& 0.36(1) 	&  0.37519399	
& 0.3607157	& 0.36266(6)	&		\\
   0.98	
&  0.357(3)	& 0.35(2) 	&  0.36517718	
& 0.340646	& 0.3422(2)	&		\\
   0.99	
&  0.355(4)	& 0.34(2) 	&  0.36170935	
& 0.33068	& 0.3319(4)	&		\\
   1.0 	
&  0.353(5)	& 0.33(3) 	&  0.35817561	
& 0.3069	& 0.307(1)	&  0.35		\\
           
&          	&         	&             
&       	&         	&  0.340(5)	\\
\end{tabular}
\end{table}
\begin{table}
\caption{
Comparison of the most accurate plaquette expansion values for
the excited state gap $ \cal G $ with those of the
{\it t}-expansion, the connected moments expansion, spin-wave
theory and series expansions, taken for various anisotropy
parameters $x$. }
\label{compare-gap-table}
\bigskip
\begin{tabular}{cccccc}
 $\cal G $    & plaquette
              & {\it t}-expansion\tablenote{Reference~\cite{woh95}}
              & CMX\tablenote{Reference~\cite{woh95}}     
              & 3rd Order\tablenote{Reference~\cite{hza92}}
              & series\tablenote{Reference~\cite{zoh91}}     \\
              & expansion 
              & D Pad\'e      
              &
              & spin-wave
              & expansion  \\
\hline
$x=0.2$	&  1.9357(6)	& 1.9338(2) 	&  1.93383753	
			& 1.942248	& 1.933815	\\
   0.5 	&  1.603(3)	& 1.594(10) 	&  1.59431537	
			& 1.629782	& 1.59736(4)	\\
   0.8 	&  0.953(9)	& 0.96(2) 	&  0.99272985	
			& 0.98798172	& 0.970(3)	\\
   0.9 	&  0.61(4)*	& 0.65(5) 	&  0.74000985	
			& 0.65721412	& 0.66(1)	\\
   0.95	&  0.50(3)*	& 0.52(10) 	&  0.60488156	
			& 0.44092456	& 0.45(3)	\\
   0.98	&  0.44(3)*	& 0.45(15) 	&  0.52114736	
			& 0.26588297	& 0.26(6)	\\
   0.99	&  0.42(5)*	& 0.4(2) 	&  0.49280432	
			& 0.18381204	& 0.17(8)	\\
   1.0 	&  0.40(5)*	& 0.3(3) 	&  0.46424870	
			& 0.0		& 0.0(1)	\\
\end{tabular}
\end{table}
\eject
\end{document}